\documentclass[12pt]{article}
\usepackage{graphicx}
\usepackage{hyperref}
\hypersetup{
 pdftitle={},%
 pdfauthor={},%
 pdfsubject={},%
 pdfkeywords={},%
 pdfstartview={},%
 bookmarksopen=true, breaklinks=true, debug=true,
 colorlinks=true, linkcolor=blue, citecolor=red, urlcolor=blue,
 hyperfigures=true
}
\def \b{{\cal B}}
\newcommand{\mbc}{M_\mathrm{BC}}

\newcommand\pubnumber{WSU--HEP--XXYY}
\newcommand\pubdate{\today}

\def\umn{
School of Physics and Astronomy\\ 
University of Minnesota, Minneapolis, Minnesota 55455, USA}

\def\Title#1{\begin{center} {\Large #1 } \end{center}}
\def\Author#1{\begin{center}{ \sc #1} \end{center}}
\def\Address#1{\begin{center}{ \it #1} \end{center}}

\newcommand\pubblock{\rightline{\begin{tabular}{l} \pubnumber\\
         \pubdate  \end{tabular}}}
\newenvironment{Abstract}{\begin{quotation}  }{\end{quotation}}
\newenvironment{Presented}{\begin{quotation} \begin{center} 
             PRESENTED AT\end{center}\bigskip 
      \begin{center}\begin{large}}{\end{large}\end{center} \end{quotation}}
\def\Acknowledgements{\bigskip  \bigskip \begin{center} \begin{large}
             \bf ACKNOWLEDGEMENTS \end{large}\end{center}}

\begin{document}
\begin{titlepage}
\pubblock

\vfill
\Title{Selected recent results on charm hadronic decays from BESIII}
\vfill
\Author{Hajime Muramatsu}
\Address{\umn}
\vfill
\begin{Abstract}
I report BESIII preliminary results on:
\begin{enumerate}
  \item Measurement of $\sigma(e^+e^-\to
    D\bar{D})$ at E$_{\rm{cm}} = 3.773$~GeV
  \item Study of the $D\bar{D}$ production line shape near E$_{\rm{cm}} = 3.773$~GeV
  \item The first observation of singly Cabibbo-suppressed decay,
    $D\to\omega\pi$
  \item Measurement of $\b(D_S^+\to\eta' X)$ and $\b(D_S^+\to\eta'\rho^+)$ .
\end{enumerate}

\end{Abstract}
\vfill
\begin{Presented}
The 7th International Workshop on Charm Physics (CHARM 2015)\\
Detroit, MI, 18-22 May, 2015
\end{Presented}
\vfill
\end{titlepage}
\def\thefootnote{\fnsymbol{footnote}}
\setcounter{footnote}{0}
%

\section{Hadronic decays of charm mesons}

Studies of hadronic decays of charm mesons
play an important role in the understanding
the weak interactions at the $c$-sector and
provide inputs for the  beauty physics.
Two of samples accumulated by the BESIII detector~\cite{bes3det}
that are taken at E$_{\rm{cm}} = 3.773$~GeV and $4.009$~GeV
are very useful to study decays of $D$ and $D_S^{\pm}$ mesons.

The former is the largest $e^+e^-$ annihilation sample in the
world to date,
$2.92$~fb$^{-1}$~\cite{bes3lumi},
that is taken around the nominal mass of $\psi(3770)$
resonance which predominantly decays into a pair of $D$ mesons.
The latter, consisting of $482$~pb$^{-1}$~\cite{dslumi}, also
produces a pair of $D_S^+D_S^-$ with a sizable production rate
($\sigma(e^+e^-\to D_S^+D_S^-) \sim~269$ pb),
providing a clean event environment to study decays of $D_S^{\pm}$.

In this proceeding, I report four preliminary measurements
from the BESIII collaboration based on the above two $e^+e^-$
annihilation data.
The first two results are studies about $D$-pair productions
at the vicinity of the $\psi(3770)$ resonance,
a measurement of observed $\sigma(e^+e^-\to D\bar{D})$ at
E$_{\rm{cm}} = 3.773$~GeV and a study of Born-level line shape of
$\sigma(e^+e^-\to D\bar{D})$. I then present the first observation
of the singly Cabibbo-suppressed decays (SCSD), $D\to\omega\pi$,
and end this report with the measurements of
$\b(D_S^+\to\eta' X)$ and $\b(D_S^+\to\eta'\rho^+)$.

\section{\boldmath $\sigma(e^+e^-\to D\bar{D})$ at E$_{\rm{cm}} = 3.773$~GeV}

Measuring observed $\sigma(e^+e^-\to D\bar{D})$ allows us to
estimate the number of $D\bar{D}$ pairs produced in our sample
by using the integrated luminosity of the
corresponding sample\cite{bes3lumi}. This can then be used
to normalize the measured signal yields to obtain a branching fraction.

As done by the CLEO collaboration~\cite{cleoxsec}, we measure the
observed cross section by a double-tag
technique, pioneered by the MARK III Collaboration~\cite{marck3}.
This takes advantage of the fact that $D$-meson production near the
$\psi(3770)$ resonance is solely through $D\bar{D}$.

Reconstructing one $D$ meson in the pair provides a single-tag yield,
$N_{ST}^i$, with a final state, $i$. 
We seek $9$ different final
states: $D^0\to (K^-\pi^+$, $K^-\pi^+\pi^0$, $K^-\pi^+\pi^+\pi^-)$,
and
$D^+\to (K^-\pi^+\pi^+$, $K^-\pi^+\pi^+\pi^0$, $K_S^0\pi^+$,
$K_S^0\pi^+\pi^0$, $K_S^0\pi^+\pi^+\pi^-$, $K^+K^-\pi^+)$.
(Unless otherwise noted, charge conjugate modes are implied throughout this
report.)
The detail reconstruction criteria can be found in 
other BESIII publications, such as Ref.~\cite{y_cp}.

$N_{ST}^i$ can be written as $N_{ST}^i = N_{D\bar{D}}\cdot\b(D\to
i)\cdot\epsilon_i$, where $N_{D\bar{D}}$ is the number of
$D\bar{D}$ produced and $\epsilon_i$ is the reconstruction
efficiency for the decay mode, $D\to i$.  Similarly, one can have
 $N_{ST}^j = N_{D\bar{D}}\cdot\b(D\to j)\cdot\epsilon_j$.
When the pair decays explicitly into two final states,
$D\to i$ and $\bar{D} \to j$, we have
$N_{DT}^{ij} = N_{D\bar{D}}\cdot\b(D\to i)\cdot\b(\bar{D}\to j)\cdot\epsilon_{ij}$.
Here, $N_{DT}^{ij}$ is the {\it double tag} yield when we
simultaneously reconstruct the two mesons in the final states of $i$
and $j$. $\epsilon_{ij}$ is the corresponding reconstruction efficiency. 
Solving these for $N_{D\bar{D}}$, one arrives at;
\[
 N_{D\bar{D}} = \frac{N_{ST}^i\cdot N_{ST}^j\cdot\epsilon_{ij}}{N_{DT}^{ij}\cdot\epsilon_i\cdot\epsilon_j}.
\]
The observed cross section is readily obtained by dividing $N_{D\bar{D}}$
by the total integrated luminosity.

\begin{figure}[htb]
\centering
\includegraphics[keepaspectratio=true,height=3.1in,angle=0]{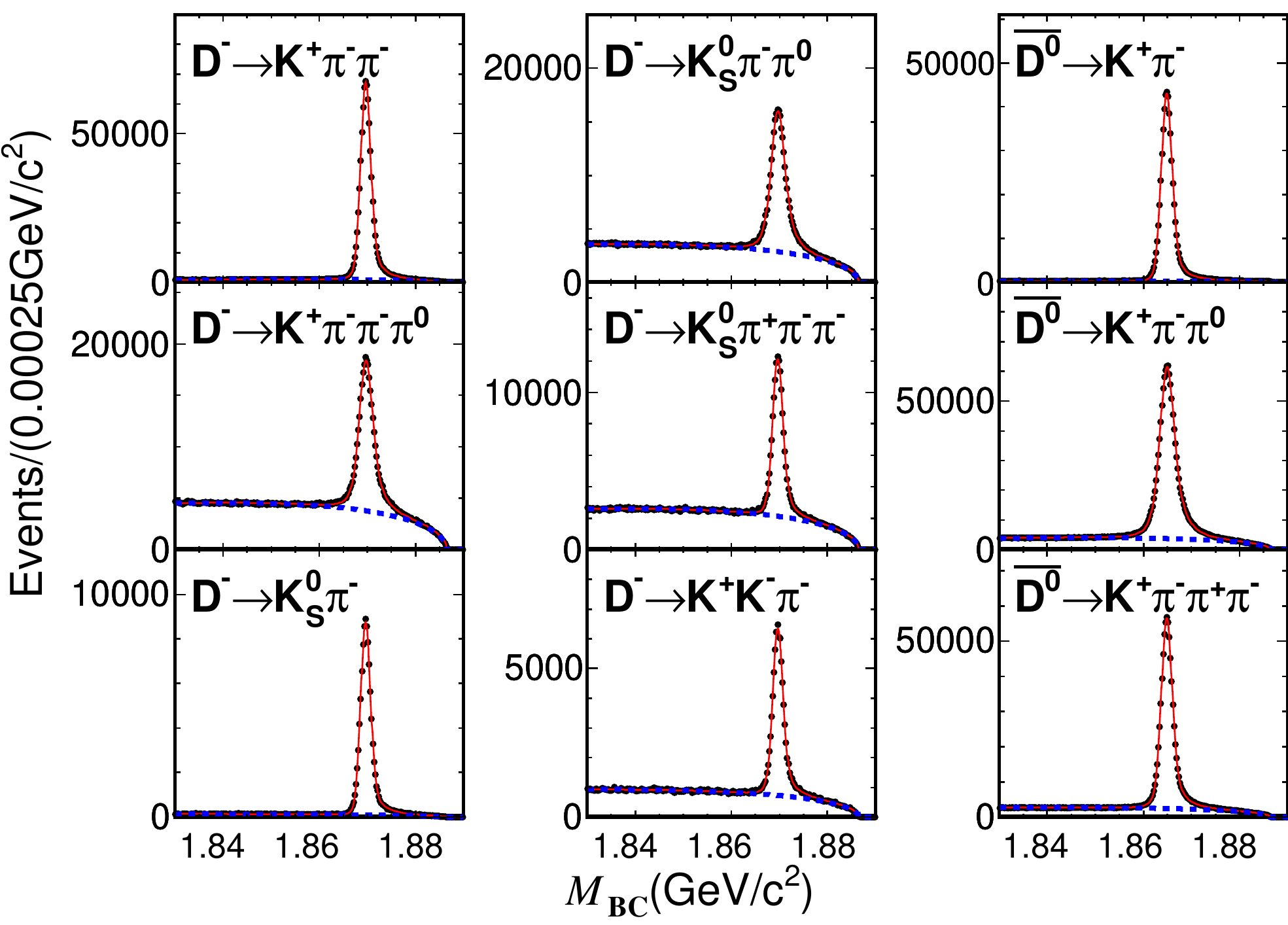}
\caption{Fits to $\mbc$ distributions of singly tagged events based on
the entire $\psi(3770)$ sample. Red curves represent the overall
fitted shapes and blue dashed curves correspond to the fitted ARGUS
background functions.}
\label{fig:DDbarST}
\end{figure}

We obtain $N_{ST}^i$ from distributions of
beam-constrained mass, $\mbc$, defined as
$\mbc  \equiv \sqrt{ E^2_{\rm{beam}} - |\vec{p}_D|^2}$.
Figure~\ref{fig:DDbarST} shows fits to $\mbc$ distributions
based on singly tagged events for the $9$ different final states.
We use a signal shape predicted by Monte Carlo (MC) simulation.
Each of these are convoluted with a Gaussian
to take into account a discrepancy in resolution between data and MC,
while using an ARGUS background function~\cite{argusBKG} to represent
the background component.

As for obtaining $N_{DT}^{ij}$, we look at a two-dimensional space,
$\mbc^{i}$ vs $\mbc^{j}$. Due to the small background of the doubly
tagged events, we simply count the yields after using
the sidebands of $\mbc$ to estimate backgrounds.

Averaging the resultant observed cross sections over different final
states ($D\to j$ and $\bar{D}\to j$ ), we have our preliminary result
shown in Table~\ref{tab:ddxsec}.
Our cross sections are consistent with the ones measured by
the CLEO collaboration~\cite{cleoxsec}. We expect our final results
to be dominated by systematic uncertainties.

\begin{table}[htb]
\begin{center}
\begin{tabular}{ccc}  
\hline\hline
 Experiment & $\sigma(e^+e^-\to D^0\bar{D^0})$ (nb) & $\sigma(e^+e^-\to D^+D^-)$ (nb) \\
\hline
This work & \multicolumn{1}{l}{$3.641\pm0.010$} 
                             & \multicolumn{1}{l}{$2.844\pm0.011$}  \\
CLEO~\cite{cleoxsec}  & \multicolumn{1}{l}{$3.607\pm0.017\pm0.056$} 
                             & \multicolumn{1}{l}{$2.882\pm0.018\pm0.042$}  \\
\hline\hline
\end{tabular}
\caption{Comparison of the measured cross sections beween
the BESIII preliminary results and the ones
  measured by the CLEO collaboration. Only statistical
  uncertainties are shown in the BESIII results.}
\label{tab:ddxsec}
\end{center}
\end{table}

\section{\boldmath Line shape of $\sigma(e^+e^-\to D\bar{D})$}

In the previous section, I report our preliminary result of
observed cross section, $\sigma(e^+e^-\to D\bar{D})$,
at  E$_{\rm{cm}} = 3.773$~GeV. 
It is of great interest to 
examine this production line shape near
the nominal mass of $\psi(3770)$ resonance.
This is done using the BESIII scan data which was taken
in $2010$, along with the main  {\it on-resonant} $\psi(3770)$
sample, in  a range of $3.642<\rm{E}_{\rm{cm}}<3.890$ GeV with the total
accumulated luminosity of $\sim 70$ pb$^{-1}$.
Such a line shape distribution allows one to extract
the $\psi(3770)$ resonance parameters.
Table~\ref{tab:shapeexp} shows some of the recent 
experimental measurements on the nominal mass of
$\psi(3770)$ resonance. There is a definite (and expected) shift
in the mass when an interference effect
is taken into account.

\begin{table}[htb]
\def\1#1#2#3{\multicolumn{#1}{#2}{#3}}
\begin{center}
\begin{tabular}{cc}  
\hline\hline
 Experiment & M$_{\psi(3770)}$ (MeV/$c^2$) \\
\hline
\multicolumn{1}{l}{BES (2008)~\cite{bes2shape}} & \multicolumn{1}{l}{$3772.0\pm1.9$} \\ 
\multicolumn{1}{l}{Belle (2008)~\cite{belleshape}} &
                                              \multicolumn{1}{l}{$3776.0\pm5.0\pm4.0$} \\ 
\multicolumn{1}{l}{{\it BABAR} (2007)~\cite{babarshape2}$\dag$} & \multicolumn{1}{l}{$3778.8\pm1.9\pm0.9$} \\ 
\multicolumn{1}{l}{{\it BABAR} (2008)~\cite{babarshape1}} & \multicolumn{1}{l}{$3775.5\pm2.4\pm0.5$} \\
\multicolumn{1}{l}{KEDR (2012)~\cite{kedr}$\dag$} & \multicolumn{1}{l}{$3779.2^{+1.8+0.5+0.3}_{-1.7-0.7-0.3}$} \\  
\hline\hline
\1{2}{l}{$\dag$ includes interference} \\
\end{tabular}
\caption{Recent experimental measurements on the mass
of the $\psi(3770)$ resonance.}
\label{tab:shapeexp}
\end{center}
\end{table}

To obtain the resonance parameters, 
we follow the procedure carried out by
the KEDR collaboration~\cite{kedr} in which we assume that there
are two sources  that produce $D\bar{D}$ final states: one from
the decay of $\psi(3770)$ and the other from non-$\psi(3770)$ decays.
To represent the non-$\psi(3770)$ decays,
we form its amplitude as a linear combination of
a constant term, which represents
the possible
contributions from higher  $c\bar{c}$
resonant states such as $\psi(4040)$, and
a Breit-Wigner form, that corresponds to
the $\psi(3686)$ tail above the $D\bar{D}$ mass threshold~\cite{DD2S}.
This approach is known as a Vector-Dominance Model (VDM), but 
we also try an exponential
form, instead of the  Breit-Wigner form,
 to see how much an alternate form affects the resultant
$\psi(3770)$ resonance parameters.

The Born-level cross section, $\sigma_{born}$, and
experimentally determined observed cross section,
$\sigma_{obs}$, are related as:
\[
\sigma_{obs}(W) = \int{z_{DD}(W\sqrt{1-x})\sigma_{born}(W\sqrt{1-x})F_{ISR}(x,W^2)}dx.
\]
\noindent Here, $z_{DD}$ is a factor for the coulomb interaction for
$D^+D^-$, $F_{ISR}(x,W^2)$ is the ISR radiator~\cite{fadin}, and
$G(W,W')$ (a Gaussian) is there to take into account the beam spread
at the initial E$_{\rm{cm}} = W$.
More details can be found in Ref.~\cite{kedr}.

\begin{figure}[htb]
\centering
\includegraphics[keepaspectratio=true,height=1.5in,angle=0]{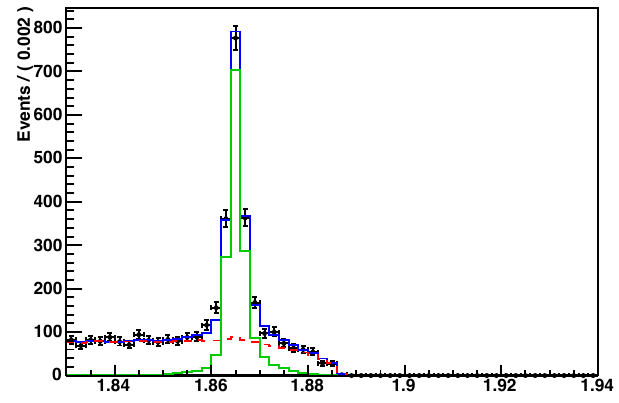}
\includegraphics[keepaspectratio=true,height=1.5in,angle=0]{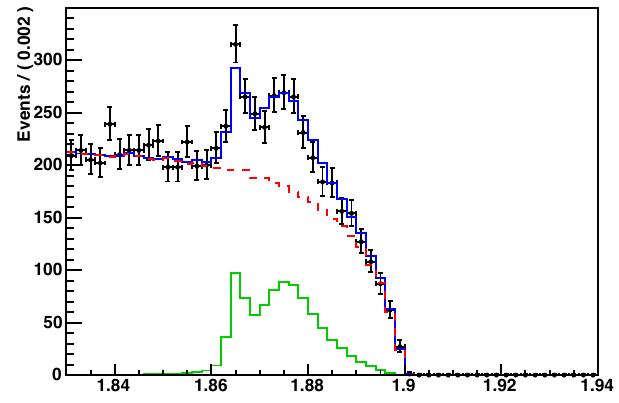}
\caption{Projections onto the $\mbc$ axes (in GeV/$c^2$) of the two-dimensional fits
($\Delta E$ vs $\mbc$) at 
E$_{\rm{cm}} \sim 3.7735$~GeV (left) and
 E$_{\rm{cm}} \sim 3.7984$~GeV (right) based on
the sum of the three $D^0$ decay modes.
 The blue histograms represent
the overall fits, while dashed red and solid green histograms
correspond to the fitted background and signal shapes, respectively.}
\label{fig:DDshapeMbc}
\end{figure}

We extract $\sigma_{born}(W)$ based on $\sigma_{obs}(W)$ with the above relation.
$\sigma_{obs}(W)$ is based on the singly tagged events
by fitting to two-dimensional space, $\Delta E$ vs $\mbc$, where
$\Delta E  \equiv E_D - E_{\rm{beam}}$ with both signal and background
shapes are fixed based on MC samples.
As an example, Fig.~\ref{fig:DDshapeMbc} shows projections
onto the $\mbc$ axes of such two-dimensional fits at 
 E$_{\rm{cm}} \sim 3.7735$~GeV (left) and
 E$_{\rm{cm}} \sim 3.7984$~GeV (right) based on
the sum of the three $D^0$ decays (see the $3^{rd}$ column of
Fig.~\ref{fig:DDbarST}).
Notice that the left plot of Fig.~\ref{fig:DDshapeMbc} peaks
at nominal mass of $D^0$, while the 
right plot of Fig.~\ref{fig:DDshapeMbc} has a $2^{nd}$ peak on the
higher side. This is due to the
larger ISR effect at this particular  E$_{\rm{cm}}$, which
our MC-based signal shape (green) reproduces quite well.

\begin{figure}[htb]
\centering
\includegraphics[keepaspectratio=true,height=2.3in,angle=0]{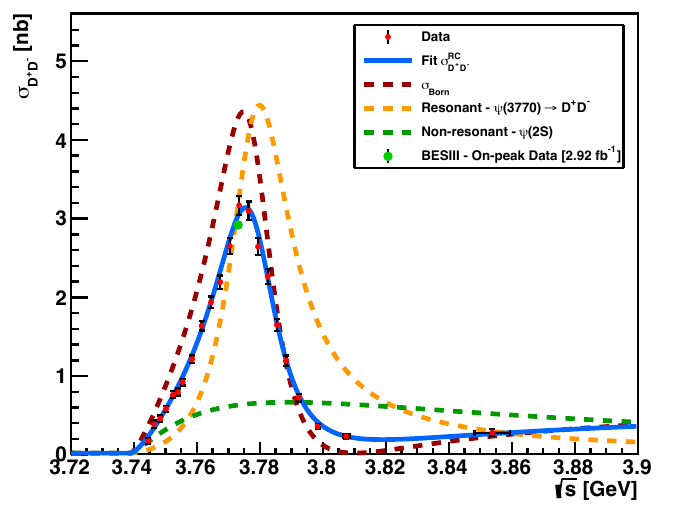}
\caption{Observed cross section, $\sigma_{obs}$ is plotted in the red
  points based on the $D^+D^-$ events in which
$D^{\pm}$ decays into the 6 different final states
(see the $1^{st}$ and the $2^{nd}$ columns of Fig.~\ref{fig:DDbarST}).
The corresponding $\sigma_{Born}$ curve is shown in dashed brown.
 }
\label{fig:DDshape}
\end{figure}

From these fits at each  E$_{\rm{cm}}$, we construct the spectrum
of the observed cross section, $\sigma_{obs}$.
As an example, we show  $\sigma_{obs}$ distribution for
the case of $D^+D^-$ (red points) in Fig~\ref{fig:DDshape}.
There, the solid blue curve is the fitted shape
to $\sigma_{obs}$, while the corresponding $\sigma_{Born}$ is
represented by the dashed brown curve.
The dashed orange and green curves are the fitted resonant
and non-resonant components (here, we use the VDM to represent the
non-resonant component).

Table~\ref{tab:ddshape} shows our preliminary results on
the nominal mass, total width, electronic partial width of
the $\psi(3770)$ resonance. The $4^{th}$ column
shows $\Gamma^{\psi(3770)}_{ee}\times\b_{D\bar{D}}$,
where $\b_{D\bar{D}} = \b(\psi(3770)\to D\bar{D})$.
This is because our fit is only sensitive to the product of the two, but
not individually.
Our preliminary result is consistent with the KEDR measurement.
In Tab.~\ref{tab:ddshape} we also show a result
based on the exponential form to represent the non-$\psi(3770)$
amplitude. As can be seen, this would likely be one of the dominant
sources of the systematic uncertainty.

\begin{table}[t]
\def\1#1#2#3{\multicolumn{#1}{#2}{#3}}
\begin{center}
\begin{tabular}{cccc}  
\hline\hline
Source & M$^{\psi(3770)}$ $($MeV/$c^2)$ & $\Gamma^{\psi(3770)}$
$($MeV$)$  & $\Gamma^{\psi(3770)}_{ee}\times\b_{D\bar{D}}$ $($eV$)$ \\
\hline
BESIII$_{\rm{VDM}}$ & $3781.5\pm0.3$ & $25.2\pm0.7$ & $230\pm18$ \\
BESIII$_{\rm{Exponential}}$ & $3783.0\pm0.3$ & $27.5\pm0.9$ &
                                                              $270\pm24$ \\
KEDR\cite{kedr} & $3779.3^{+1.8}_{-1.7}$ & $25.3^{+4.4}_{-3.9}$ &
  $160^{+78}_{-58}$, $420^{+72}_{-80}$ (a)\\
PDG\cite{pdg14} & $3773.\pm0.3$ & $27.2\pm1.0$ & 
  $[262\pm18]\times\b_{D\bar{D}}$ \\
\hline\hline
\1{4}{l}{(a) Two solutions were obtained from their fit.} \\
\end{tabular}
\caption{BESIII preliminary results based on the two different
forms of the non-$\psi(3770)$ amplitudes, VDM ($\psi(3686)$)
and an exponential shape, are shown, along with
the result from the KEDR collaboration as well as the current PDG value.
In the $4^{th}$ column, $\b_{D\bar{D}} = \b(\psi(3770)\to D\bar{D})$.
}
\label{tab:ddshape}
\end{center}
\end{table}


\section{\boldmath $D \to \omega \pi$}

For Cabibbo-suppressed charm decays, such as
the yet to be observed SCSD $D\to\omega\pi$,
measurements are difficult due to
low signal statistics and high backgrounds.
For the case of $D\to\omega\pi$,
the most recent experimental search was carried out
by the CLEO collaboration~\cite{cleoomgpi}. They
set upper limits,
$\b(D^+\to\omega\pi^+) < 3.0\times 10^{-4}$ and
$\b(D^0\to\omega\pi^0) < 2.6\times 10^{-4}$ at $90\%$
confidence level (C.L.).
In the mean time, H.~Y. Cheng and C.~W. Chiang predict
the $\b(D\to\omega\pi)$ could be at an order of $1\times10^{-4}$~\cite{omgpipred}.

We start with reconstructing one of the $D\bar{D}$ pairs with the same
9 final states (see Fig.~\ref{fig:DDbarST}). Then in the other $D$
decay, we look for $D^{+(0)}\to\omega\pi^{+(0)}$, where
$\omega\to\pi^+\pi^-\pi^0$ and $\pi^0\to\gamma\gamma$.
To improve the signal-to-noise ratio, we also select
a certain range on the helicity-like angle of $\omega$,
$\theta_{\rm{helicity}}$, which is defined as
an opening angle between the direction of the normal to
the $\omega\to\pi^+\pi^-\pi^0$ plane and the direction of the
parent $D$ meson in the $\omega$ rest frame.
We require
$|H_\omega| = |\cos{\theta_{\rm{helicity}}}| > 0.54 (0.51)$ for
$D^+$ ($D^0$) that are optimized based on a MC study.

\begin{figure}[htb]
\centering
\includegraphics[keepaspectratio=true,height=2.3in,angle=0]{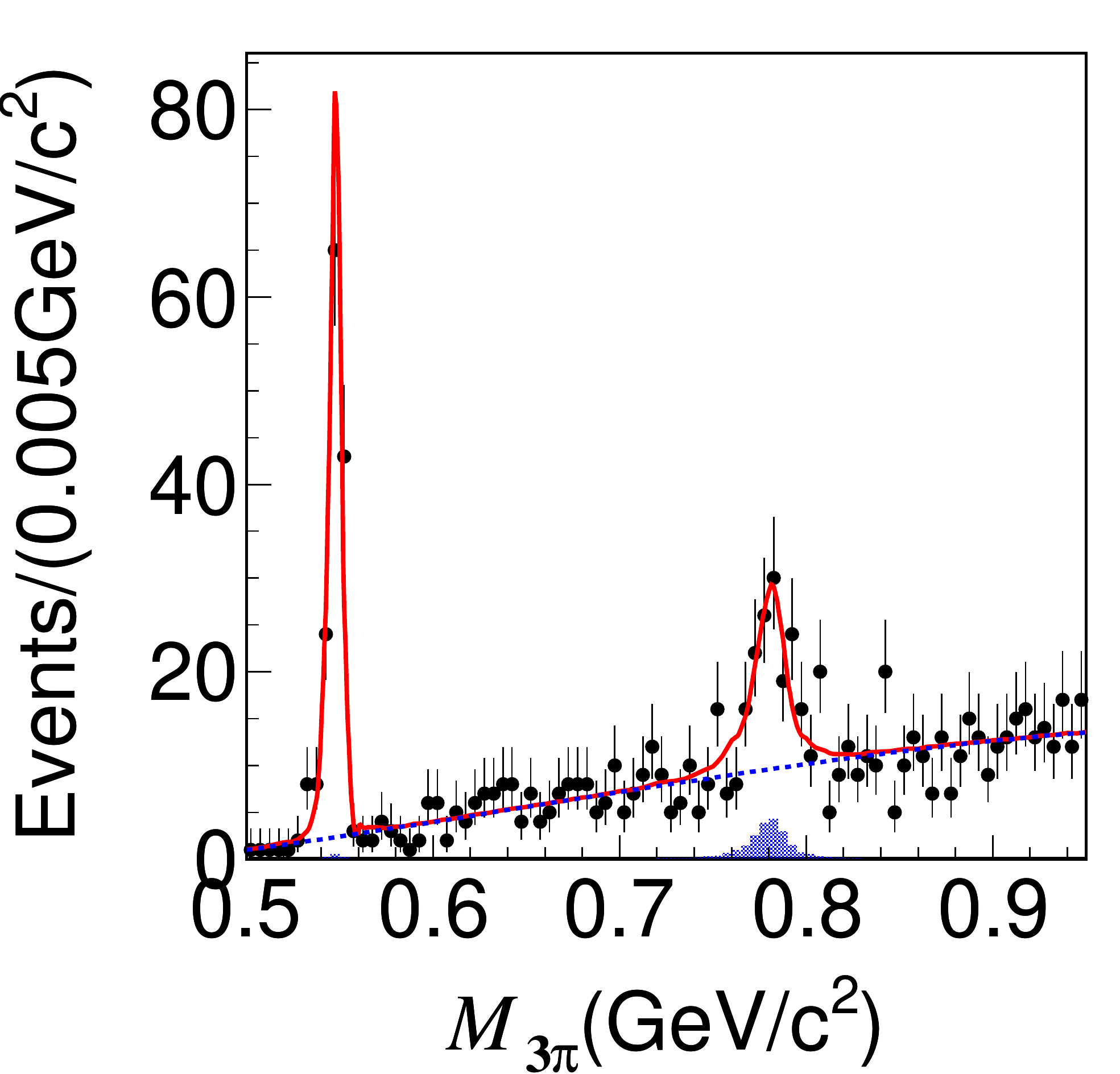}
\includegraphics[keepaspectratio=true,height=2.3in,angle=0]{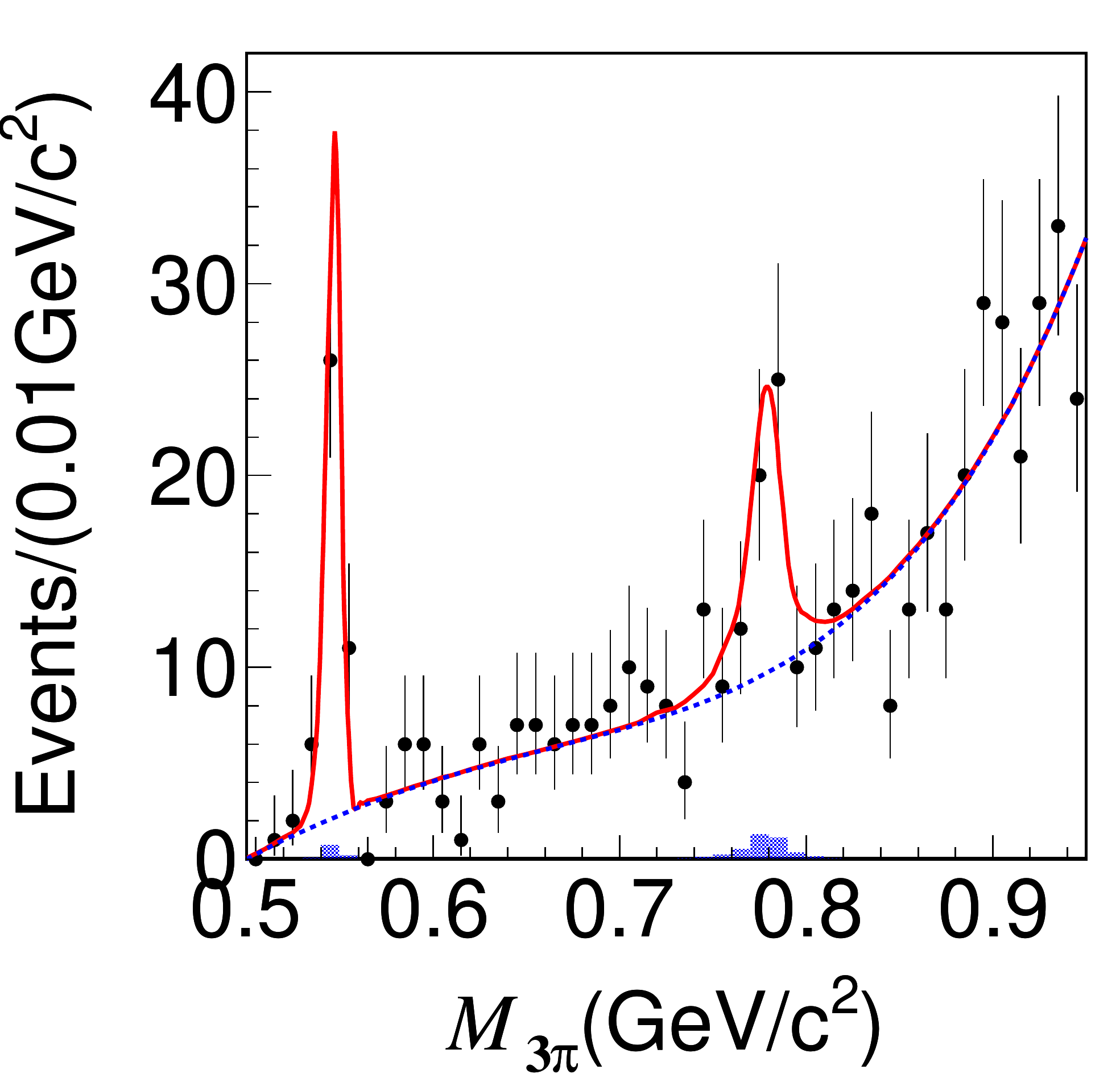}
\caption{Distributions of invariant mass of $\omega\to\pi^+\pi^-\pi^0$
  for $D^+\to\pi^+\pi^-\pi^0\pi^+$ (left) and
$D^0\to\pi^+\pi^-\pi^0\pi^0$. 
The solid red lines are the overall fits,
while the dashed blue lines represent the fitted polynomials
The filled histograms represent the peaking backgrounds, estimated
by the sidebands of $\mbc$ distributions.}
\label{fig:omgpi_omg}
\end{figure}

With additional requirements on $\mbc$ and $\Delta E$
to be consistent with a $D\bar{D}$ pair production,
we extract our signal yields by fitting to the distributions
of invariant mass of $\omega\to\pi^+\pi^-\pi^0$ as shown
in Fig.~\ref{fig:omgpi_omg}. We use MC-based signal shapes,
along with polynomials to represent their background shapes.
Figure~\ref{fig:omgpi_omg} also shows the expected peaking
backgrounds (represented by filled histograms) which are
estimated by the sidebands of $\mbc$ distributions.
The extracted signal yields correspond to
a statistical significance of $5.4\sigma (4.1\sigma)$ 
for $D^+ (D^0) \to\omega\pi^+(\pi^0)$,
respectively.

\begin{figure}[htb]
\centering
\includegraphics[keepaspectratio=true,height=1.5in,angle=0]{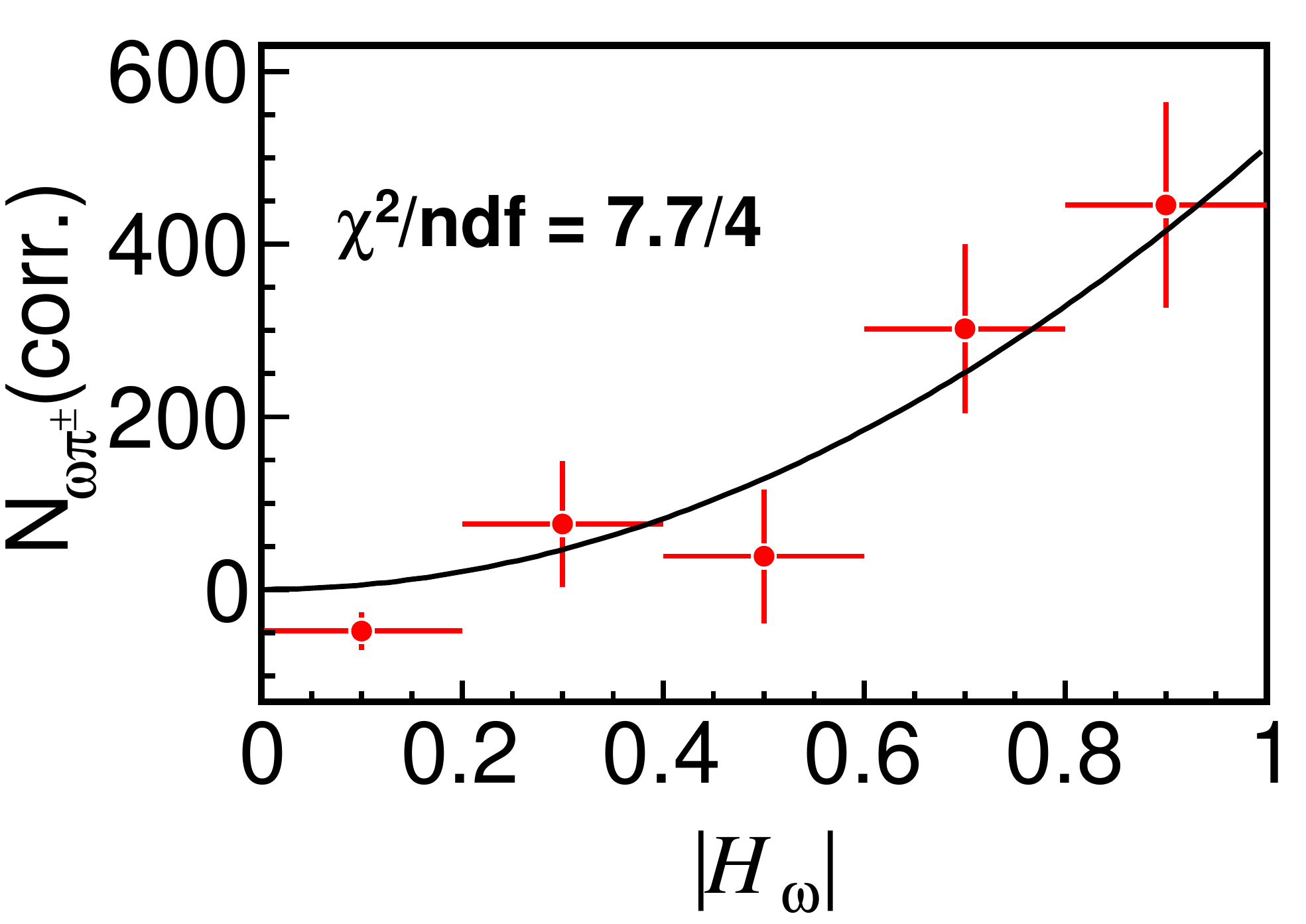}
\includegraphics[keepaspectratio=true,height=1.5in,angle=0]{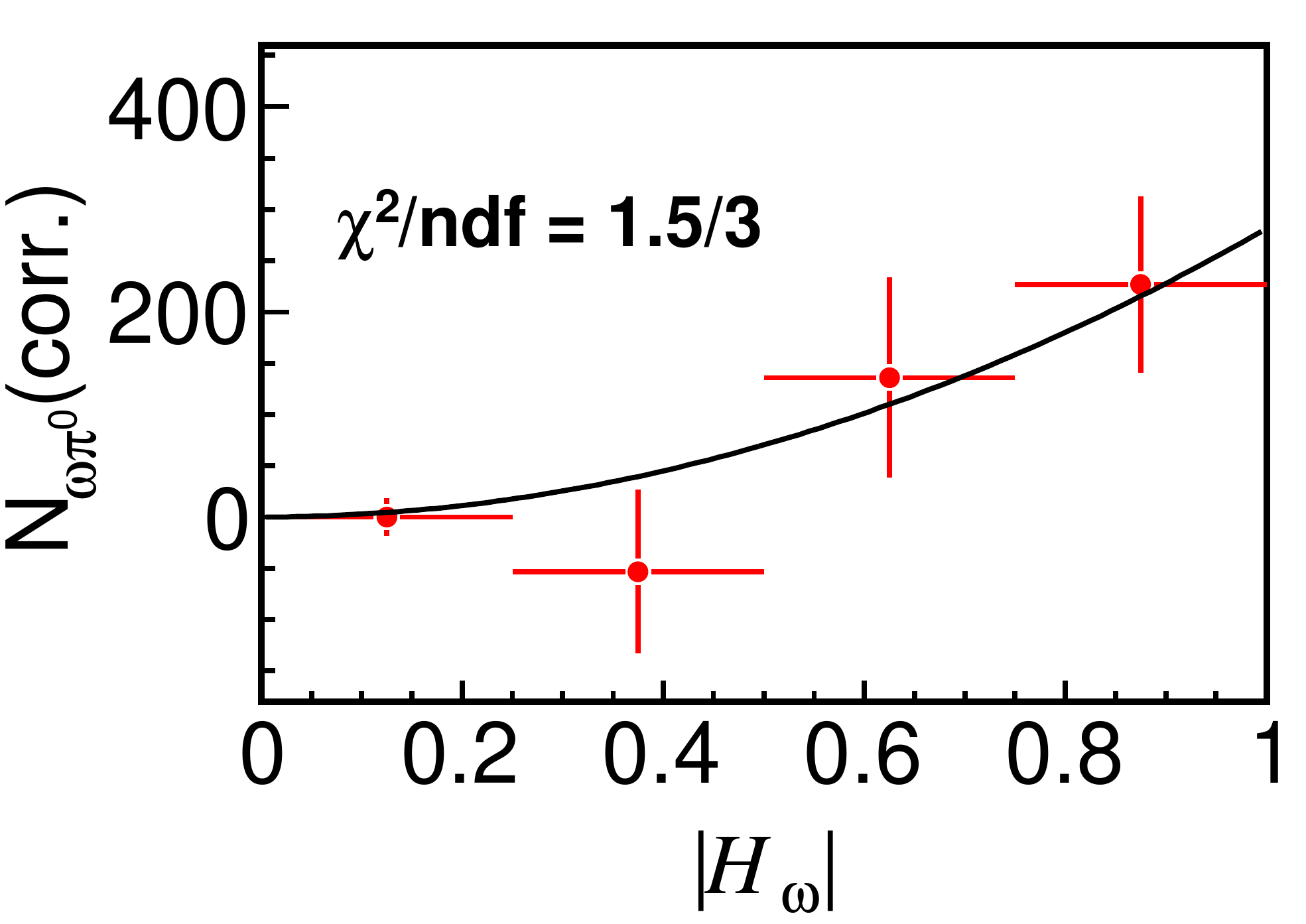}
\caption{Efficiency-corrected signal yields in the $|H_\omega|$ bins
for candidates of $D^+\to\omega\pi^+$ (left) and
$D^0\to\omega\pi^0$ (right). 
The black lines are the fitted quadratic shapes.}
\label{fig:omgpi_omg_hel}
\end{figure}

We also check to see if the $D\to\omega\pi$ candidates produce
the expected distribution of the helicity angle.
Figure~\ref{fig:omgpi_omg_hel} shows the distributions of $|H_\omega|$
in which we can see the expected $H^2_\omega = \cos^2\theta_{\rm{helicity}}$.

\begin{figure}[htb]
\centering
\includegraphics[keepaspectratio=true,height=2.3in,angle=0]{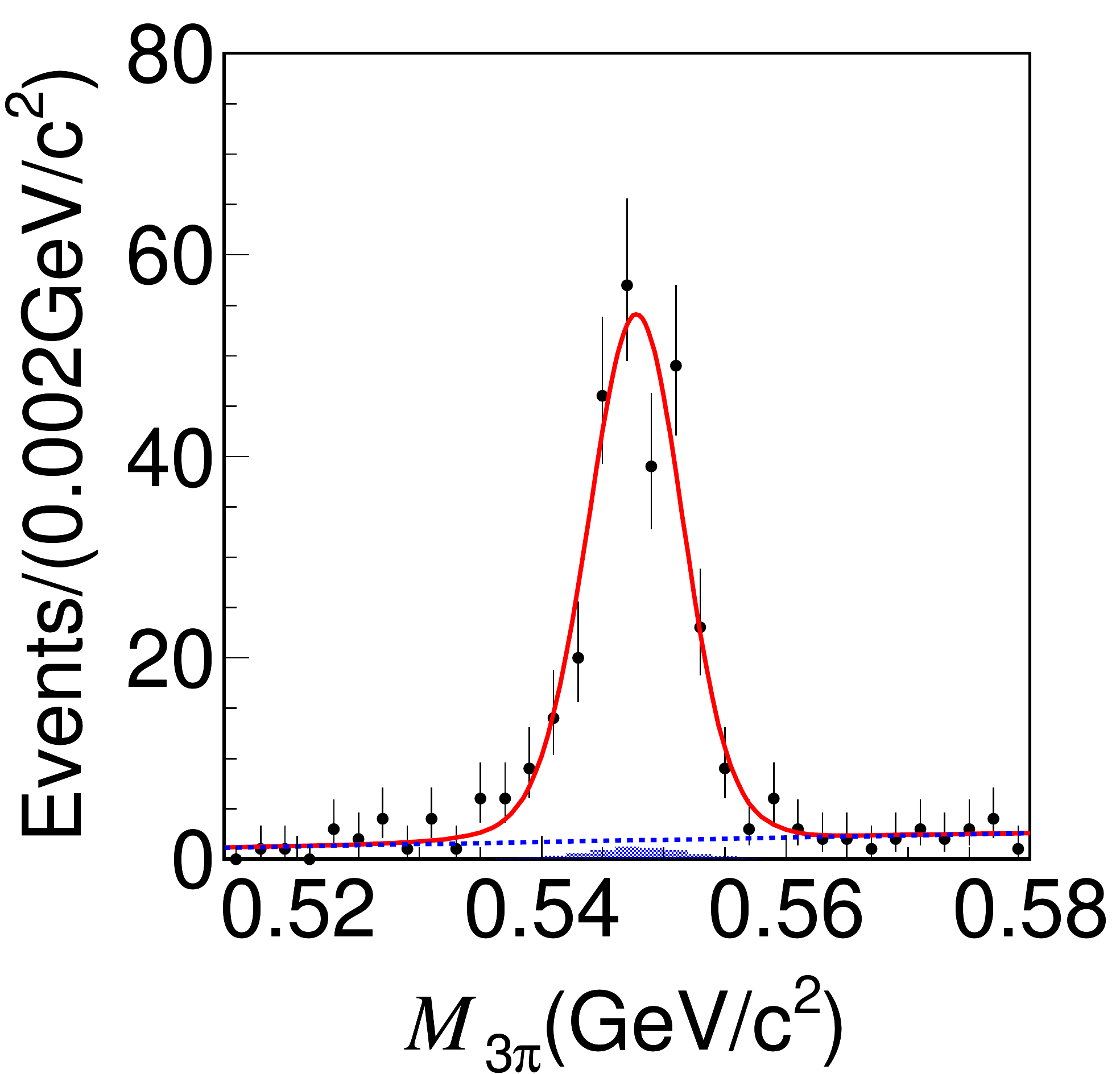}
\includegraphics[keepaspectratio=true,height=2.3in,angle=0]{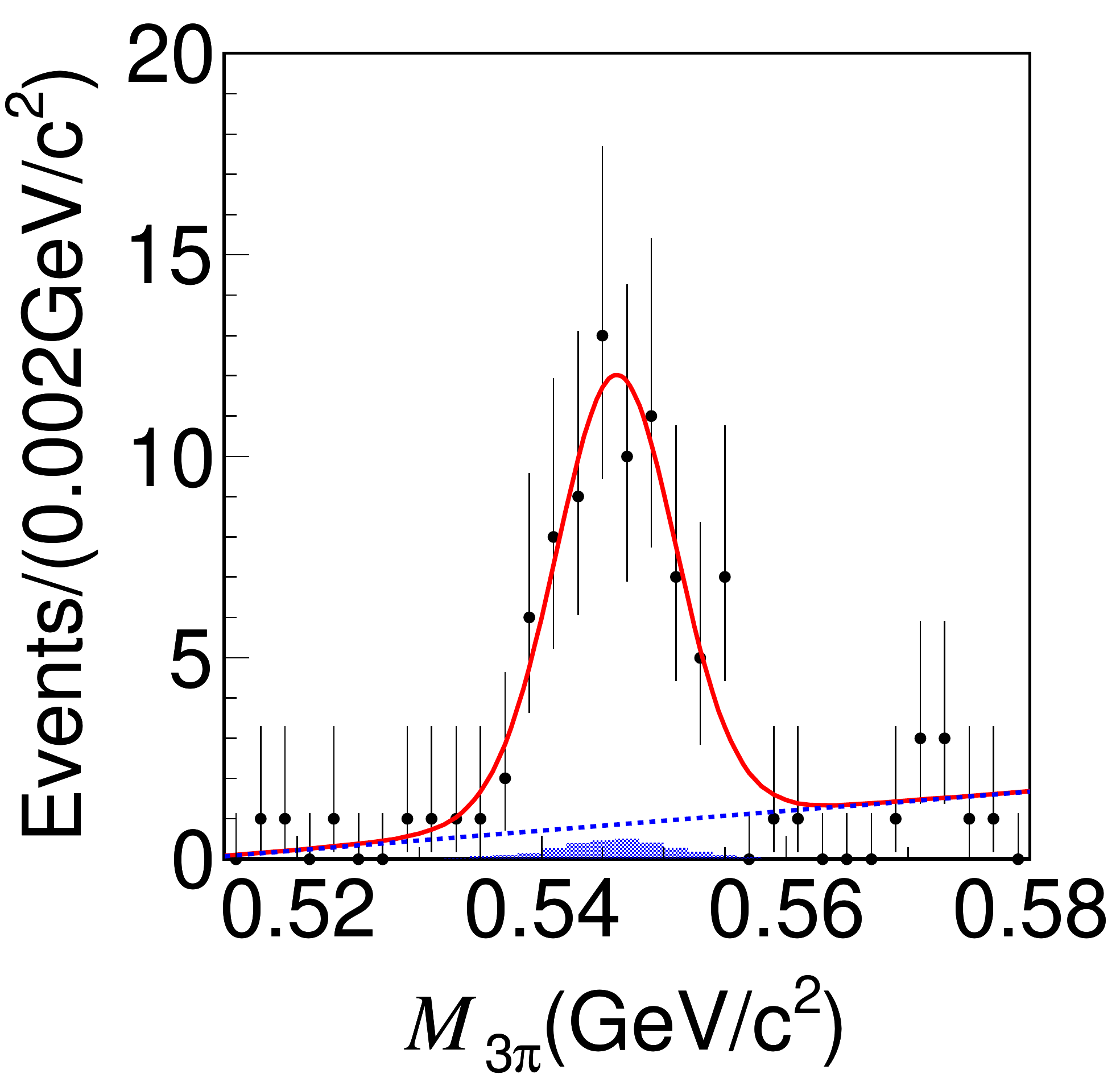}
\caption{Fits to distributions of invariant mass of
  $\omega\to\pi^+\pi^-\pi^0$
for the candidates of $D^+\to\eta\pi^+$ (left)
and $D^0\to\eta\pi^0$ (right).
The filled histograms represent the peaking backgrounds which are
estimated by the sideband regions of both signal and tag sides
of $\mbc$ distributions.}
\label{fig:omgpi_eta}
\end{figure}

In Fig.~\ref{fig:omgpi_omg}, we can also see peaks that correspond to
$D\to\eta\pi$ candidates. We extract these candidates by fitting to
the same invariant mass distributions of $\omega\to\pi^+\pi^-\pi^0$
with much narrower fit ranges, and without the requirement
on the $|H_\omega|$.
Figure~\ref{fig:omgpi_eta} shows such fits from which we also
measure $\b(D\to\eta\pi)$.

Table~\ref{tab:omgpi} 
shows our preliminary branching fraction measurements.
The measured $\b(D\to\eta\pi)$ are consistent with the known
values~\cite{pdg14},
while $\b(D\to\omega\pi)$ are measured for the first time.

\begin{table}[htb]
\begin{center}
\begin{tabular}{ccc}  
\hline\hline
 Decay mode & This work & PDG value\cite{pdg14} \\
\hline
$D^+\to\omega\pi^+$ & $(2.74\pm0.58\pm0.17)\times10^{-4}$ &  $<3.4\times10^{-4}$ at $90\%$ C.L. \\
$D^0\to\omega\pi^0$ & $(1.05\pm0.41\pm0.09)\times10^{-4}$ &  $<2.6\times10^{-4}$ at $90\%$ C.L. \\
$D^+\to\eta\pi^+$ & $(3.13\pm0.22\pm0.19)\times10^{-4}$ &  $(3.53\pm0.21)\times10^{-3}$ \\
$D^0\to\eta\pi^0$ & $(0.67\pm0.10\pm0.05)\times10^{-4}$ &  $(0.68\times0.07)\times10^{-3}$  \\
\hline\hline
\end{tabular}
\caption{Preliminary result on the measured $\b(D\to\omega\pi)$.}
\label{tab:omgpi}
\end{center}
\end{table}

\section{\boldmath $D_S^+ \to \eta X$ and $D_S^+ \to \eta\rho^+$}

The situation of $\b(D_S^+\to\eta'\rho^+)$ is rather interesting.
If we sum the all known exclusive rates with $\eta'$ in $D_S^+$ decays
 in the PDG~\cite{pdg14}, we arrive at $(18.6\pm2.3)\%$, while
$\b(D_S^+\to\eta' X) = (11.7\pm1.7)\%$~\cite{cleoetapincl}.
Among the $D_S^+$ decays that involve $\eta'$,
the largest single exclusive rate is
$\b(D_S^+\to\eta'\rho^+) = (12.5\pm2.2)\%$~\cite{cleo2etaprho}.
However, a recent measurement is about a half of it, 
$\b(D_S^+\to\eta'\pi^+\pi^0) =
(5.6\pm0.5\pm0.6)\%$~\cite{cleocetaprho}
which appears to solve the inconsistency mentioned above.
B.~Bhattacharya and J.~L.~Rosner come up with
two predictions, 
$\b(D_S^+\to\eta'\rho^+) = (2.9\pm0.3)\%$ and
$(1.89\pm0.20)\%$~\cite{etaprhopred}, while
F.~S. Yu  {\it et al.} predict
$(3.0\pm0.5)\%$~\cite{etaprhopred2}
by factorization methods.

\begin{figure}[htb]
\centering
\includegraphics[keepaspectratio=true,height=3.3in,angle=0]{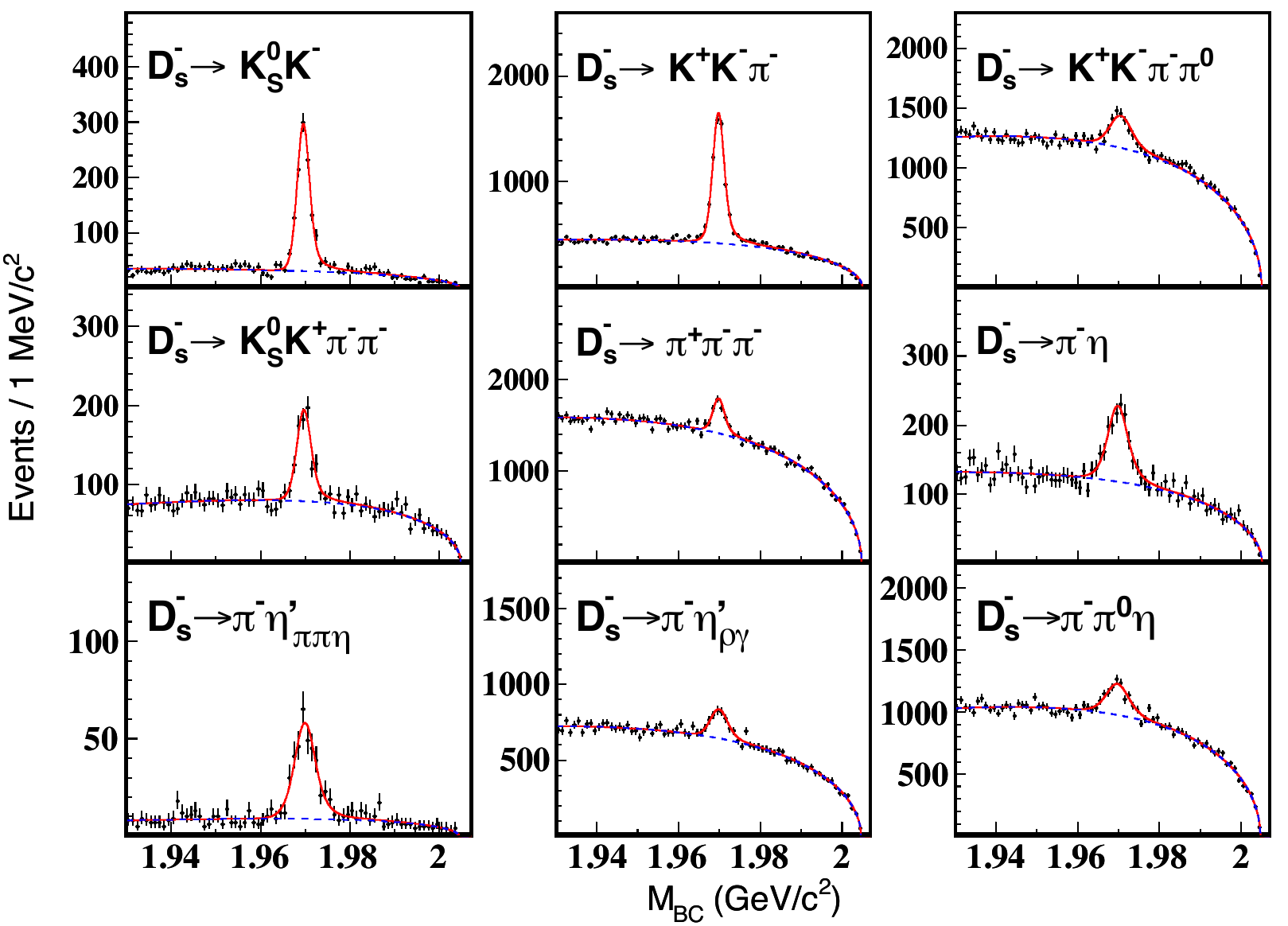}
\caption{Fits to $\mbc$ distributions of the selected $9$ different
  final states of $D_S^+$ decays. The red curves correspond to
the total fits, while the blue dashed curves represent the fitted
background shapes by the ARGUS background functions~\cite{argusBKG}.}
\label{fig:ds_st}
\end{figure}

We can use our sample taken at E$_{\rm{cm}} = 4.009$~GeV to measure
these branching fractions to confirm the recent measurement.
At this energy, the $D_S^{\pm}$ is produced in a pair.
To measure the inclusive rate, $D_S^+\to\eta' X$, we employ
a double-tag technique in which we reconstruct
its tag side in $9$ decay modes shown in Fig.~\ref{fig:ds_st}.
From these $\mbc$ distributions, the single-tag yields are readily
obtained.

\begin{figure}[htb]
\centering
\includegraphics[keepaspectratio=true,height=2.3in,angle=0]{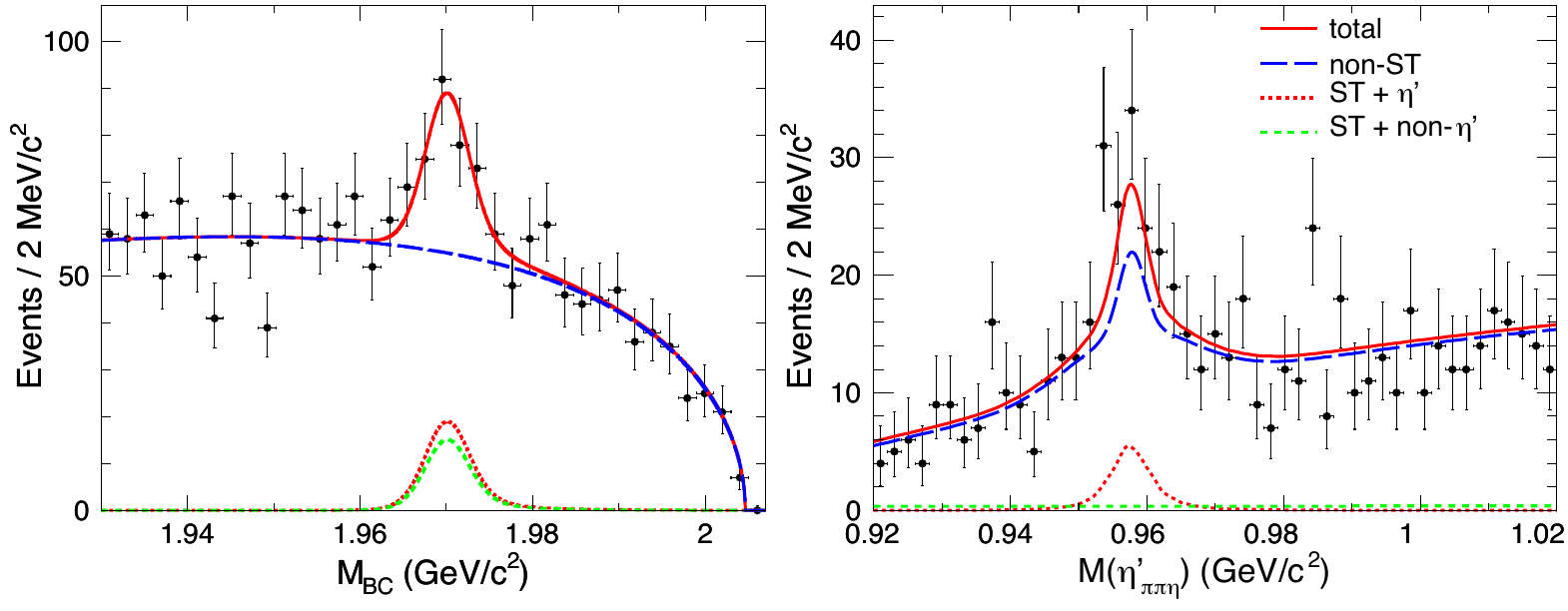}
\caption{Fit to two-dimensional space, $M_{\pi^+\pi^-\eta}$ vs
$\mbc$, where $\mbc$ is the tag side of the beam-constrained mass.
Shown here are the fitted result, projected onto the
 $\mbc$ axis (left) and onto the $M_{\pi^+\pi^-\eta}$ axis (right).
Solid red curves correspond to the overall fit, while dashed blue and green
curves are fitted peaking backgrounds.}
\label{fig:ds_etapincl}
\end{figure}

To obtain the double-tag yields, we reconstruct
the $9$ final states of $D_S^+$ decays and look for
the other $D_S^{\mp}$ decays in the final states with
$\eta'\to\pi^+\pi^-\eta(\to\gamma\gamma)$
based on the remaining particles. 
If there is more than one $\eta'$ candidate, we choose the one that
gives the minimum $|M_{\pi^+\pi^-\eta} - M_{\eta'}(PDG)|$.
We fit to a two-dimensional space, $M_{\pi^+\pi^-\eta}$ vs
$\mbc$, to extract the signal yields, where $\mbc$ is the
tag side of the beam-constrained mass.
Figure~\ref{fig:ds_etapincl} shows such fits, projected
onto the $\mbc$ axis (left) and onto the
$M_{\pi^+\pi^-\eta}$ axis (right). We use MC-based distributions
to represent the signal shape. As for the background shapes,
an ARGUS background function~\cite{argusBKG} is used
on the $\mbc$ direction, while 
the smooth and peaking backgrounds 
on the $M_{\pi^+\pi^-\eta}$ axis
are represented by
a polynomial plus double Gaussian shapes.

From this fit, $68\pm14$ events are observed as signal candidates.
This translates into $\b(D_S^+\to\eta' X) = (8.8\pm1.8\pm0.5)\%$
which agrees with the known value~\cite{pdg14}.

To measure $\b(D_S^+\to\eta'\rho^+)$, we simply use the single-tag
method by reconstructing 
$D_S^+\to\eta'\rho^+$, where $\rho^+\to\pi^+\pi^0$.
We require the reconstructed $\eta'$ mass to be within
$3\sigma$ of the known mass~\cite{pdg14}, the invariant mass
$M_{\pi^+\pi^0}$ be within $0.17$~GeV/$c^2$ of the known
$\rho$ mass~\cite{pdg14}, and finally its $\Delta E$ be consistent
with zero.

The signal yield is extracted by fitting to two-dimensional  space,
$\mbc$ vs $\cos\theta_{\pi^+}$, where $\theta_{\pi^+}$
is the helicity angle of the $\pi^+$ from the $\rho$ decay.
We expect to see $\cos^2\theta_{\pi^+}$ for $D_S^+\to\eta'\rho^+$, while
$D_S^+\to\eta'\pi^+\pi^0$ events should be independent of
$\theta_{\pi^+}$.

\begin{figure}[htb]
\centering
\includegraphics[keepaspectratio=true,height=2.3in,angle=0]{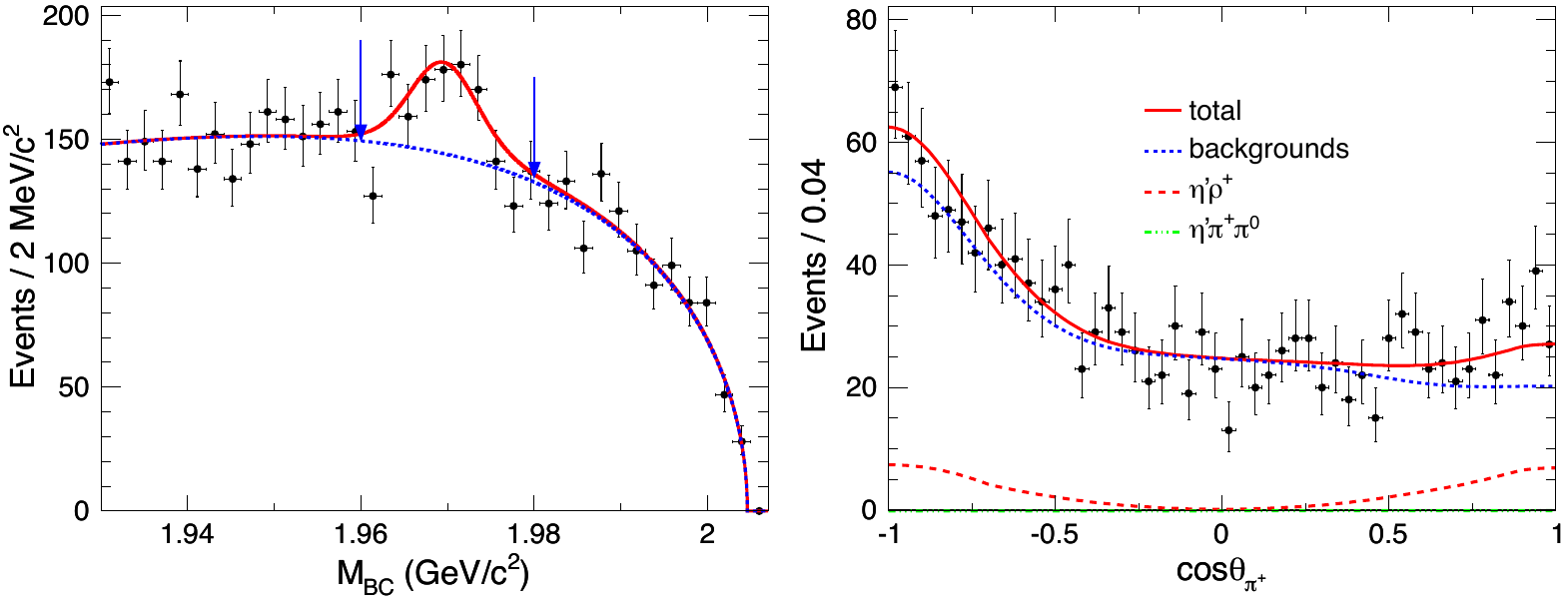}
\caption{Projections onto the $\mbc$ axis (left) and the
$\cos\theta_{\pi^+}$ axis with an additional requirement
of $(1.960<\mbc<1.980)$~GeV/$c^2$ (right) of the two-dimensional fit.}
\label{fig:ds_etaprho}
\end{figure}

Figure~\ref{fig:ds_etaprho} shows projections onto
the $\mbc$ axis (left) of such two dimensional fit. On the right, a projection onto the
$\cos\theta_{\pi^+}$ axis with an additional requirement
of $(1.960<\mbc<1.980)$~GeV/$c^2$ is shown.
Signal shapes are based on MC simulation.
To represent the background shapes,
an ARGUS background function~\cite{argusBKG} is used
on the $\mbc$ axis,
while a fixed non-$D_S^+$ background shape is employed
on the $\cos\theta_{\pi^+}$ axis, estimated from the $\mbc$ sidebands.

The fit yields
$210\pm50$ and $-13\pm56$ events
for $D_S^+\to\eta'\rho^+$ and
$D_S^+\to\eta'\pi^+\pi^0$ candidates, respectively.
We normalize the rate by $D_S^+\to K^+K^-\pi^+$ mode to obtain
$\b(D_S^+\to\eta'\rho^+)/\b(D_S^+\to K^+K^-\pi^+) =
1.04\pm0.25\pm0.07$.
Or with the known $\b(D_S^+\to K^+K^-\pi^+)$~\cite{pdg14},
we arrive at $\b(D_S^+\to\eta'\rho^+) = (5.8\pm1.4\pm0.4)\%$
which confirms the recent measurement by the
CLEO collaboration~\cite{cleocetaprho}.
We also set an upper limit on the non-resonant decay,
$\b(D_S^+\to\eta'\pi^+\pi^0) < 5.1\%$ at $90\%$ C.L.

\section{\boldmath Conclusion}

Four preliminary results on the hadronic final states
in the decays of $D$ and $D_S^{\pm}$ mesons based on the two
recent BESIII samples are reported. The measurements based on the
world's largest $e^+e^-$ annihilation sample taken
at E$_{\rm{cm}} = 3.773$~GeV provide statistically superior results
than the previous experimental results, while the study of
decays of $D_S^{\pm}$ based on
the sample at E$_{\rm{cm}} = 4.009$~GeV shows the very clean
event environment at BESIII. It would be very exciting to pursue
our $D_S$ program as the collaboration plans to take a few fb$^{-1}$
of $e^+e^-$ annihilation sample at
 E$_{\rm{cm}} = 4.180$~GeV  in $2015-2016$, where
the production rate of $D_S^{\pm}$ is much higher,
$\sigma(e^+e^-\to D_S^{*\pm}D_S^{\mp}) \sim 900$~pb.

\Acknowledgements
I would like to thank Derrick Toth, Andy Julin, Xiaoshuai Qin, and
Peilian Liu for preparing and providing the figures and comments.


\end{document}